\documentclass[a4paper,conference]{IEEEtran}
\usepackage[left=1.57cm,right=1.57cm,top=0.95cm,bottom=2.54cm]{geometry}
\IEEEoverridecommandlockouts
\usepackage{cite}
\usepackage{amsmath,amssymb,amsfonts}
\usepackage{balance}
\usepackage{subfigure}
\usepackage{algorithmic}
\usepackage{graphicx}
\usepackage{textcomp}
\usepackage{xcolor}
\usepackage{comment}

\usepackage{array}

\begin{document}

\title{Zigbee vs. Matter over Thread: Understanding IoT Protocol Performance in Practice}
\author{\IEEEauthorblockN{Massimo Nobile, Fabio Palmese, Antonio Boiano, Alessandro E. C. Redondi, Matteo Cesana}
\IEEEauthorblockA{\textit{DEIB, Politecnico di Milano} \\
\textit{Milan, Italy}\\
\{name.surname\}@polimi.it}
}

\maketitle

\begin{abstract}
The widespread adoption of the Internet of Things (IoT) has positioned smart homes as paradigmatic examples of distributed automation systems, where reliability, efficiency, and interoperability depend critically on the underlying communication protocol. Among the low-power wireless technologies available for this scenario, Zigbee and Matter over Thread have emerged as leading contenders. While Zigbee represents a mature, non-IP mesh networking solution, Matter over Thread introduces an IP-based architecture designed to unify device interoperability across different ecosystems. However, despite extensive documentation of their design principles, there is a lack of empirical, comparative performance data under realistic network conditions.
This paper presents a comprehensive experimental comparison between the two protocols, conducted on a testbed built from commercially available hardware. The proposed methodology focuses on different key performance dimensions, such as scalability, responsiveness, and fault tolerance. 
The results reveal that Zigbee achieves a lower baseline overhead and faster route recovery, making it more responsive in static small-scale deployments. Matter over Thread, conversely, exhibits superior scalability and robustness, maintaining stable throughput and predictable latency across multi-hop scenarios.
Overall, we highlight that Zigbee and Matter over Thread embody distinct trade-offs between agility, efficiency, and scalability.

\end{abstract}

\begin{IEEEkeywords}
Matter, Zigbee, IoT
\end{IEEEkeywords}

\section{Introduction}
\label{sec:intro}

The Internet of Things (IoT) has become a foundational pillar of modern automation and control engineering, with smart homes representing one of its most mature and rapidly evolving application domains. Contemporary smart home environments consist of complex distributed systems composed of numerous heterogeneous devices, sensors, actuators, and controllers, that must communicate reliably, efficiently, and securely under stringent energy and latency constraints. The overall performance and robustness of such systems are fundamentally determined by the characteristics of the underlying wireless communication protocols, which directly affect critical engineering dimensions including latency, throughput, scalability, fault tolerance, and energy efficiency. Consequently, the selection of the communication protocol is a central design decision that strongly influences the reliability, responsiveness, and long-term maintainability of a smart home ecosystem.

In residential deployments, communication technologies must satisfy requirements that differ substantially from those of traditional data networks. Devices are often battery-powered, expected to operate unattended for long periods, and deployed in dynamic topologies where nodes may frequently join, leave, or fail. For these reasons, low-power wireless mesh networking protocols have emerged as the dominant solution for sensing, actuation, and control tasks in smart homes. Among these, Zigbee has achieved widespread adoption over the past decade and remains deeply embedded in commercially successful ecosystems, including lighting systems such as Philips Hue, smart home devices from IKEA, and a broad range of sensors and actuators integrated into major automation platforms. More recently, Matter over Thread has emerged as a transformative standard, backed by a large industry consortium, with the goal of addressing long-standing interoperability challenges by defining a common application layer that operates over an IPv6-based, low-power Thread mesh network. As a result, Zigbee and Matter over Thread increasingly represent the two dominant paradigms for infrastructure-grade, low-power communication in smart home environments.

Alternative connectivity technologies are also present in consumer IoT deployments. Wi-Fi–based solutions are commonly employed for bandwidth-intensive devices such as cameras, multimedia systems, and voice assistants. However, Wi-Fi typically entails higher energy consumption, increased contention on the home network, and limited scalability when supporting large numbers of constrained devices. Moreover, Wi-Fi deployments are typically based on star topologies and lack the native, low-power, self-healing mesh networking mechanisms required for pervasive sensing and actuation, making consumer-grade Wi-Fi less suitable as a primary communication substrate in this context. Bluetooth Low Energy (BLE), including BLE Mesh, plays a significant role in IoT ecosystems as well, particularly for device provisioning, proximity-based control, and smartphone-centric interactions. While BLE Mesh enables many-to-many communication, it relies on managed flooding rather than structured routing and is generally optimized for sporadic traffic patterns and limited network sizes. Consequently, BLE-based solutions are less commonly employed as the backbone for always-on, infrastructure-level smart home automation involving persistent device availability, multi-hop routing, and large-scale operations such as coordinated firmware updates.  

Accordingly, this work restricts its scope to routing-based, low-power mesh protocols designed for persistent smart home infrastructures, focusing on Zigbee and Matter over Thread as representative and widely adopted technologies in this space.

Despite the growing relevance of Matter over Thread and the long-standing deployment of Zigbee, comprehensive empirical comparisons of their real-world performance remain scarce. Existing studies often focus on architectural differences, theoretical capabilities, or isolated performance metrics, while systematic experimental evaluations under equivalent conditions are limited. This lack of comparative analysis makes it difficult for system designers and practitioners to make informed decisions when selecting a protocol for scalable and reliable smart home deployments.

This work addresses this gap by conducting a rigorous experimental comparison of Zigbee and Matter over Thread within a controlled smart home testbed. The two protocols are evaluated under equivalent hardware and network conditions to answer the following research questions:
\begin{itemize}
    \item \textit{Scalability and Stability}: Which protocol maintains greater stability and responsiveness as the number of connected devices and network depth increase?
    \item \textit{Responsiveness and Efficiency}: Which protocol provides superior responsiveness in user-facing operations (e.g., remote device activation) and more efficiently handles demanding tasks such as group firmware updates?
    \item \textit{Fault Tolerance}: In the presence of device or node failures, which protocol minimizes disruption and ensures network continuity without manual intervention?
\end{itemize}

To answer these questions, this work collects labeled datasets of Matter over Thread and Zigbee traffic and presents a data-driven comparative analysis of the two protocols. We design and deploy an experimental testbed using identical hardware for both technologies and define a methodology that enables a comprehensive evaluation of scalability, efficiency, and fault tolerance in realistic smart home scenarios.

The experimental results reveal distinct trade-offs between the two ecosystems. Zigbee exhibits lower baseline overhead and faster route recovery, making it well suited for small to medium-sized and relatively static networks. In contrast, Matter over Thread demonstrates superior scalability and more stable performance under heavily loaded and growing network conditions. These findings highlight that the choice between the two protocols is not one of absolute superiority, but rather a matter of engineering trade-offs between agility and stability, as well as efficiency and scalability.

The remainder of this paper is organized as follows. Section~\ref{sec:background} provides an overview of the architectural differences between Zigbee and Matter over Thread. Section~\ref{sec:related} reviews related work on IoT communication protocol performance evaluation. Section~\ref{sec:testbed} describes the experimental testbed and methodology. Section~\ref{sec:performance} presents and discusses the comparative performance results. Finally, Section~\ref{sec:conclusion} concludes the paper and outlines directions for future research.

\section{Background}
\label{sec:background}

The Zigbee and Matter over Thread protocol stacks differ substantially in their network, transport, and application-layer design.
Zigbee, first standardized in 2005 \cite{ZigbeeSpec}, is a mature and vertically integrated wireless communication stack. In contrast, Thread, standardized in 2015 \cite{ThreadSpec}, is a more recent, IP-based networking protocol designed with flexibility and interoperability as primary objectives.

\begin{figure}
    \centering
    \includegraphics[width=0.7\linewidth]{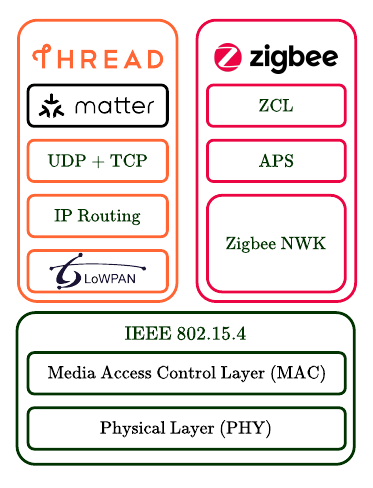}
    \caption{Diagram of the two analyzed protocols stacks}
    \label{fig:stack}
\end{figure}

Despite these architectural differences, the stack diagram of the protocols in Figure \ref{fig:stack} illustrates that both technologies rely on an identical foundation in the Physical (PHY) and Medium Access Control (MAC) layers, as defined by the IEEE~802.15.4 standard \cite{2020IEEENetworks}. Operating in the 2.4\,GHz ISM band at a raw data rate of 250\,kbps, both Zigbee and Thread are optimized for low-power, battery-operated devices. As a result, any performance disparities observed between the two systems cannot be attributed to radio hardware characteristics, but rather to the design choices and behavior of the upper protocol layers.

\subsection{Network Layer and Routing Strategy}

Zigbee employs a proprietary network layer based on non-IP, 16-bit short addressing. Its architecture enforces a hierarchical topology centered around a single Coordinator, which is responsible for Personal Area Network (PAN) formation, security key distribution, and bridging to external networks \cite{SiliconLabs2021UG103.02:Fundamentals}. Routing in Zigbee is based on an AODV-derived reactive protocol: when a node lacks a route to a destination, it initiates a network-wide Route Request (RREQ) broadcast, which propagates until the destination is reached and a return path is established. While this approach enables rapid route discovery and often yields minimum-hop paths, it can lead to significant control overhead and broadcast storms in large, dense, or unstable networks, particularly during topology changes.

Thread adopts a fundamentally different approach. The mesh network is decoupled from external IP connectivity through the use of Thread Border Routers (TBRs), and multiple Border Routers can coexist within the same network to provide redundancy and seamless failover \cite{ThreadGroup2022ThreadPaper}. Routing is handled proactively by router-capable nodes using the Mesh Link Establishment (MLE) protocol. Thread routers periodically exchange routing advertisements containing link quality metrics, such as Link Margin, which are used to compute a cumulative Path Cost. This metric-based routing strategy allows Thread to favor more stable multi-hop paths over shorter but unreliable ones, prioritizing network robustness and consistency over the immediate reactivity characteristic of Zigbee’s on-demand routing.

\subsection{Adaptation and Transport Layers}

A further distinction affecting efficiency and reliability emerges at the adaptation and transport layers. Zigbee encapsulates transport functionality within its proprietary Application Support Sub-layer (APS), which provides lightweight acknowledgements and retransmission mechanisms tailored specifically to the Zigbee stack.

Thread, as an IP-native protocol, must accommodate the relatively large IPv6 headers within the constrained frame size of IEEE~802.15.4. This challenge is addressed through the 6LoWPAN (IPv6 over Low-Power Wireless Personal Area Networks) adaptation layer, which performs aggressive header compression, reducing the standard 40-byte IPv6 header to as little as 6–7 bytes, and manages packet fragmentation and reassembly \cite{Mulligan2007TheArchitecture}. By exposing standard transport protocols to the upper layers, Thread enables the use of User Datagram Protocol (UDP) and Transmission Control Protocol (TCP). In practice, most application traffic, including Matter commands, relies on UDP due to its low latency and minimal overhead. TCP support is retained for scenarios requiring guaranteed delivery, such as Over-the-Air (OTA) firmware updates, where the additional control overhead is justified by the need for reliability.

\subsection{Application Layer}

The application layer ultimately determines how devices interpret messages and interact within a smart home ecosystem. Zigbee defines its own proprietary application framework, the Zigbee Cluster Library (ZCL), later known as Dotdot, which standardizes device models and command semantics but confines interoperability to the Zigbee ecosystem.

Thread itself is application-layer agnostic and intentionally does not define this layer, allowing any IP-based application protocol to operate on top of the network. This design choice enables the deployment of Matter, released in 2022 \cite{MatterSpec}, as a unified application layer spanning multiple physical and link-layer technologies, including Thread, Wi-Fi, and Ethernet. Matter adopts a data model conceptually derived from ZCL, retaining notions such as clusters and attributes, 
while decoupling the application logic from the underlying radio technology. This architecture enables end-to-end interoperability across heterogeneous vendor ecosystems without the proprietary translation gateways that have historically fragmented smart home deployments.

\subsection{Architectural Implications for Performance and Resilience}

The architectural differences outlined above have direct implications for the performance and fault-tolerance characteristics of Zigbee and Matter over Thread deployments. Zigbee’s reactive, broadcast-based routing strategy enables rapid path discovery but may incur increased control traffic and recovery delays during topology changes, particularly in dense or unstable networks. Its reliance on a single Coordinator further introduces a potential point of failure that can affect network-wide availability.

In contrast, Thread’s proactive routing mechanism, combined with metric-based path selection and support for multiple Border Routers, is explicitly designed to enhance network stability and resilience. The use of 6LoWPAN enables IP interoperability at the cost of additional adaptation-layer processing, while standardized transport protocols introduce well-defined trade-offs between latency and reliability. At the application layer, Matter’s IP-native design prioritizes interoperability and uniform device behavior across heterogeneous ecosystems, while introducing additional protocol overhead compared to vertically integrated stacks.

Taken together, these design choices suggest inherent trade-offs between routing overhead, recovery behavior, latency, and robustness that are not observable at the PHY or MAC layers. Quantifying the practical impact of these trade-offs under realistic deployment conditions requires systematic experimental evaluation, which motivates the analysis presented in this work.

\section{Related Work}
\label{sec:related}

Considering the very recent standardization and commercial rollout of the Matter protocol, the current body of literature remains limited in both scope and experimental depth. Existing works primarily focus on architectural descriptions, interoperability goals, and security considerations, while empirical performance evaluations on realistic hardware deployments are still scarce. In particular, studies assessing Matter over Thread in comparison with established smart-home protocols remain largely absent. In the following, we review the most relevant contributions related to Matter, Thread, and comparable low-power wireless technologies.

Among the few works explicitly addressing Matter over Thread, Madadi-Barough et al. \cite{Madadi-Barough2025Matter:Homes} present one of the earliest experimental evaluations using a hardware testbed. Their study analyzes the encapsulation overhead introduced by the Matter protocol stack when operating over Thread and Wi-Fi, and reports latency measurements at the application layer. While this work provides valuable insight into protocol layering costs, it does not include a comparative analysis with alternative smart-home communication standards such as Zigbee or Bluetooth Low Energy (BLE), thereby limiting its usefulness for protocol selection and system-level design decisions.

A broader set of publications investigates the performance characteristics of the Thread protocol independently of the Matter application layer. These studies typically focus on latency, reliability, and scalability in mesh networking scenarios. For instance, \cite{NXPSemiconductors2017ThreadNetwork} evaluates unicast and multicast latencies in large Thread networks using CoAP at the application layer, highlighting the influence of network depth and traffic patterns. Khattak et al. \cite{Khattak2023PerformanceCities} analyze the effects of jitter and packet loss under varying payload sizes and bitrates, providing insight into Thread’s behavior under constrained conditions. Similarly, Sistu et al. \cite{Sistu2019PerformanceSystems} examine unicast and multicast communication in lighting control applications, with particular emphasis on network coverage time and node reception rates. A complementary analytical perspective is presented in \cite{Lan2019LatencyThread}, where a system-level latency model of the Thread stack is proposed and validated experimentally by isolating the contributions of individual protocol layers.

Real-world deployment aspects are addressed in \cite{Grohmann2021InterferenceEvaluation}, where the authors implement a Thread testbed in a densely populated office environment to evaluate network performance under realistic interference conditions. Their results demonstrate a measurable reduction in link losses when interference is minimized, underlining the importance of environmental factors in mesh-based IoT networks. The resilience and overhead of Thread networks are further investigated in \cite{Mihaeljans2023OpenthreadAnalysis} using the OpenThread Network Simulator, where packet overhead, reachability, and robustness against router failures are evaluated under controlled traffic injection and node disconnection scenarios.

In the wider context of IoT protocol evaluation, several comparative studies have analyzed low-power wireless technologies from both theoretical and practical perspectives. Works such as \cite{Rzepecki2019IoTSP:Study} and \cite{SiliconLabs2019AN1142:Comparison} compare protocols including Zigbee, Thread, and BLE in terms of throughput, latency, energy efficiency, and scalability. Bluetooth Low Energy, in particular, is often highlighted for its widespread adoption and low power consumption; however, its reliance on star or scatternet topologies and its limited native support for large-scale, self-healing mesh networks constrain its applicability to highly distributed smart-home automation scenarios. Crucially, none of these comparative studies considers Matter at the application layer, nor do they evaluate the interaction between Matter and its underlying transport technologies in realistic deployments.

Finally, with respect to network resilience and fault tolerance, methodological references are available outside the wireless IoT domain. Prytz et al. \cite{Prytz2007NetworkTopology} propose a topology-based testing methodology for evaluating recovery times in wired Ethernet networks, which closely parallels the experimental approach adopted in this work. For Zigbee specifically, documented measurements of network self-healing times are reported in \cite{Ayurzana2024BuildingProtocol}, providing a valuable reference baseline for validating fault-tolerance results.

In summary, while Thread has been extensively analyzed as a networking protocol and Matter has been studied from architectural and security perspectives, there remains a clear lack of experimental studies that (i) evaluate Matter over Thread in realistic hardware testbeds and (ii) directly compare its performance and resilience against mature alternatives such as Zigbee. Addressing this gap is the primary objective of the present work.
\begin{figure}[t]
    \centering
    \includegraphics[width=1\linewidth]{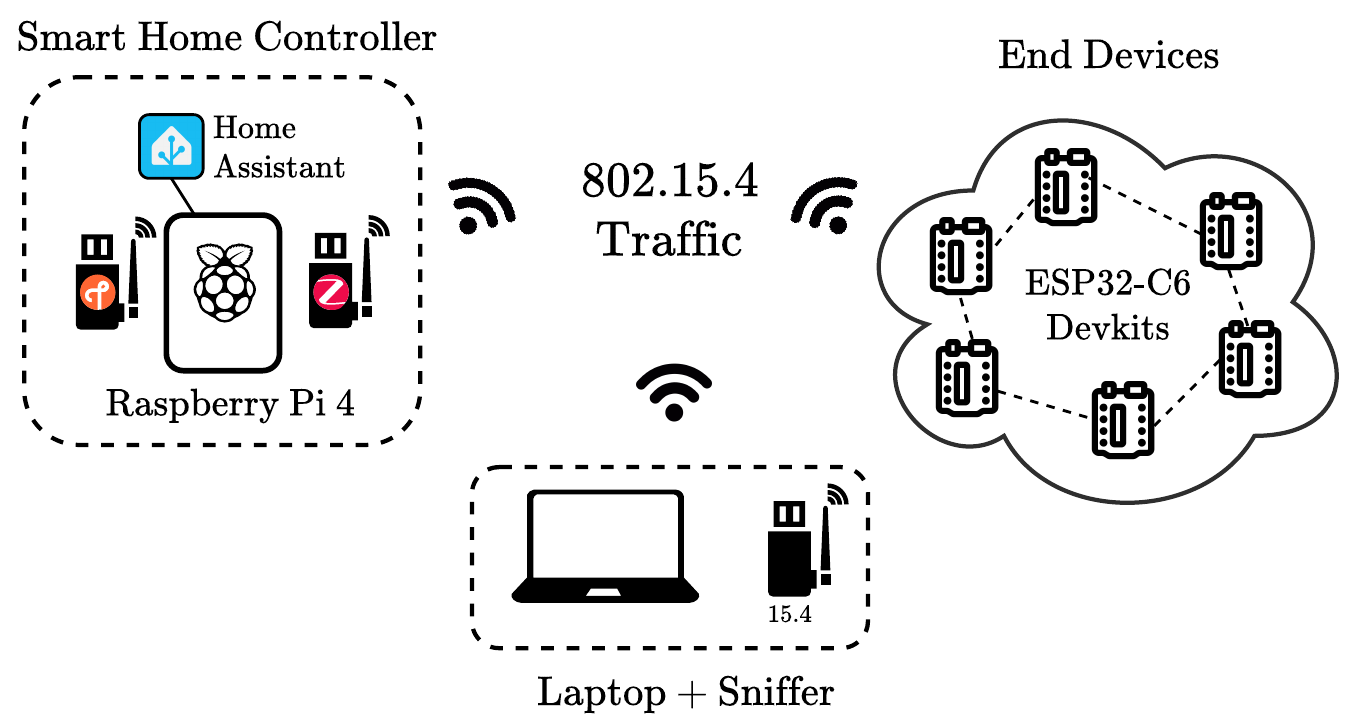}
    \caption{Sketch of the testbed architecture}
    \label{fig:diagram}
\end{figure}
\section{Testbed Setup}
\label{sec:testbed}

\begin{figure}
    \centering
    \subfigure[Full mesh topology]{\includegraphics[width=0.24\textwidth]{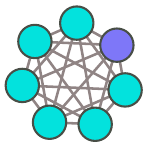}\label{fig:fullmesh}}
    \subfigure[Chain topology]{\includegraphics[width=0.24\textwidth]{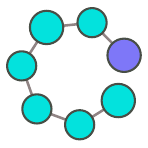}\label{fig:chain}}
    \caption{\centering Diagrams of the implemented topologies\linebreak (Generated by the OTBR web GUI)}
    \label{fig:topo}
\end{figure}

This work relies on a versatile and reproducible hardware testbed composed entirely of commercially available components. The devices were selected for their broad availability, strong community support, and compatibility with the evaluated protocols. All experiments were conducted on this platform to ensure fairness, reproducibility, and consistent environmental conditions.
Figure \ref{fig:diagram} reports a sketch of the main elements composing the testbed. The architecture consists of a central smart home controller acting as network coordinator for Zigbee and as a border router for Thread, a set of ESP32-C6 development boards serving as mesh nodes, and a dedicated passive packet sniffer. This configuration enables the construction of diverse topologies (such as those represented in Figure \ref{fig:topo}) and facilitates comprehensive observation of network behavior across both link and application layers.
The smart home environment is controlled by a Raspberry Pi 4 4GB running Home Assistant OS, an operating system specifically built around \textit{Home Assistant}, an open-source home automation software\footnote{https://www.home-assistant.io/}.
To interact with Matter devices in Home Assistant, the Matter controller is implemented as a standalone process with the \textit{Matter Server} add-on, to which the integration connects using a WebSocket connection.
A similar implementation is used to form or join a Thread network and make Home Assistant a Thread Border Router, with the \textit{OpenThread Border Router} add-on and its corresponding integration, using an IEEE 802.15.4 capable radio. For this purpose, we selected the Sonoff ZBDongle-E (based on the Silicon Labs EFR32MG21 chip),  flashing it with the OpenThread RCP firmware v2.4.5.0 using the \textit{Universal Silicon Labs Flasher}.\\
For what concerns Zigbee, the network is created and managed through the \textit{ZHA} integration. The radio used for this task is a Texas Instruments CC2531 USB dongle, flashed with the Z-Stack coordinator firmware (1.2 HA version), using a \textit{CC-DEBUGGER} programmer and the \textit{SmartRF Flash-Programmer} software.\\
The end devices consist of six ESP32-C6 boards selected for their dual support of IEEE 802.15.4 (enabling Thread and Zigbee) and Wi-Fi/BLE (supporting Matter commissioning over BLE), making it a versatile platform for modern IoT development. This ensures protocol parity across identical hardware, eliminating variability from device-specific factors. Each board runs custom firmware derived from Espressif ESP-IDF SDK examples, emulating typical smart home devices while exposing command-line interfaces for diagnostics and measurement.\\
Low-level protocol monitoring is achieved through a passive packet sniffing subsystem, designed to capture IEEE 802.15.4 frames non-intrusively. The sniffer comprises a laptop running Ubuntu 24.04 LTS and a CC2531 USB dongle flashed with \textit{Packet-Sniffer} firmware from Texas Instruments. The firmware enables promiscuous capture mode, forwarding all received frames to the host over USB without being part of the network.
Captured data are processed using the \textit{whsniff} utility, which streams raw frames to PCAP format for later analysis in Wireshark. Wireshark provides detailed visualization, filtering, and decoding of the captured traffic. A custom Matter protocol dissector was integrated to enable deep inspection of Matter-over-Thread application-layer exchanges, facilitating direct comparison with Zigbee communication patterns.
\section{Performance Analysis}
\label{sec:performance}
This section details the experimental methodology for the tests and presents the quantitative results obtained. The main goal is to provide a direct, data-driven comparison of the Matter-over-Thread and Zigbee protocol stacks across three key performance areas: protocol overhead, end-to-end latency/throughput, and mesh network resilience. To cover the different key aspects, we define three different methodological experiments to be executed on the previously defined testbed, as thoroughly covered in the following subsections.

\subsection{Overhead and Scalability}
\label{sec:overhead}

Protocol overhead represents the amount of network traffic required to sustain basic network operation, excluding any user-generated application traffic. It includes control messages such as beacons, keep-alives, neighbor advertisements, routing control packets, and retransmissions. Quantifying this overhead is essential to assess how efficiently a protocol scales as the network grows in size and complexity.

\subsubsection{Experimental Methodology}

To obtain a representative comparison, protocol overhead was evaluated under different network topologies, node counts, and activity conditions. For each protocol, a dedicated firmware was developed using official examples provided within the Espressif ESP-IDF framework for ESP32 devices. The Zigbee firmware was derived from the \textit{HA\_color\_dimmable\_light} example included in ESP-ZIGBEE-SDK version 1.6.5 and configured to operate as a router device. The Matter over Thread firmware was built from the \textit{light} example included in ESP-MATTER version 1.4 and configured as a router-capable Full Thread Device (FTD).

In both cases, the firmware emulated the behavior of a standard smart bulb and was commissioned and paired with the same smart home controller (a Raspberry Pi), ensuring that application-level behavior was functionally equivalent across the two ecosystems.

Two network topologies, illustrated in Figure~\ref{fig:topo}, were considered to reflect common smart home deployments. A fully connected mesh topology represents an idealized single-room scenario in which all devices are within mutual radio range. A chain topology represents multi-hop layouts such as corridors or elongated floor plans, where nodes are forced to forward traffic along a linear path. The chain topology was enforced through MAC address filtering at the firmware level, ensuring deterministic forwarding paths and enabling reliable packet capture using a single sniffer.

\subsubsection{Traffic Conditions and Metrics}

To separate baseline protocol overhead from traffic generated by user interactions, experiments were conducted under two activity conditions. In the \textit{Idle} state, no application commands were issued and only background maintenance traffic, such as beacons, keep-alives, and routing updates, was observed. In the \textit{Controlled Traffic} state, an automated script issued an On/Off toggle command every five seconds. In the chain topology, commands were always directed to the final node in the chain to maximize the number of hops, while in the fully connected mesh topology the destination node was randomly selected at each interval.

To study scalability, all experiments were repeated for increasing network sizes ranging from one to six nodes. For each unique configuration, a 15-minute packet capture was recorded. As primary metrics, both the total packet rate and the application-layer packet rate were considered.

To distinguish application-layer traffic from protocol overhead, packets were classified offline based on protocol headers and message types, using Wireshark filters on the obtained packet captures \texttt{.pcapng} files.
Application-layer packets were defined as messages carrying user-level control commands exchanged between the controller and end devices. For Zigbee, these correspond to packets containing APS (Application Support Sublayer) layer (\texttt{zbee\_aps}), while for Matter over Thread they correspond to packets containing Matter application payloads (\texttt{matter}), carried over UDP/IPv6. All remaining packets, including MAC and network-layer control frames, routing messages and neighbor advertisements, were classified as protocol overhead. This classification procedure was applied consistently across all scenarios and protocols.

\subsubsection{Results}

\begin{figure}[t]
    \centering
    \includegraphics[width=1\linewidth]{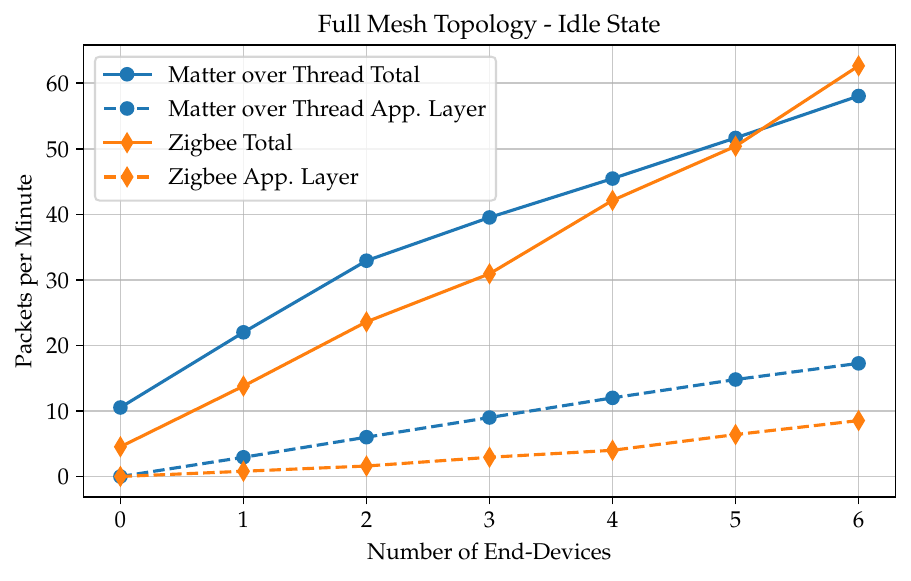}
    \caption{Packet rate for the overhead test on a fully connected mesh topology under idle conditions.}
    \label{fig:FullMeshNoLoad}
\end{figure}

\begin{figure}[t]
    \centering
    \includegraphics[width=1\linewidth]{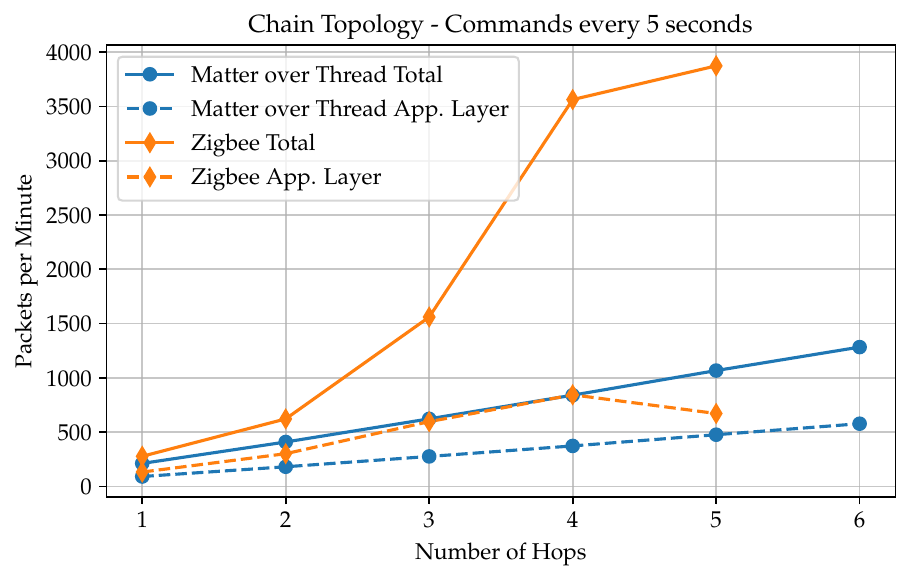}
    \caption{Packet rate for the overhead test on a chain topology under controlled traffic conditions.}
    \label{fig:ChainLoad}
\end{figure}

The results reveal a clear divergence in scalability and stability between Zigbee and Matter over Thread across both topological and traffic conditions.

In the fully connected mesh topology under idle conditions (Figure~\ref{fig:FullMeshNoLoad}), Zigbee initially exhibits lower baseline traffic for small network sizes, generating less overhead than Matter over Thread for up to five nodes. However, as the network grows, Zigbee’s overhead increases at a steeper rate and surpasses that of Matter over Thread at six nodes. This increase is not driven by application-layer traffic, as Zigbee consistently generates fewer application-layer packets than Matter over Thread in this scenario. This behavior suggests that Zigbee’s internal network maintenance mechanisms, such as neighbor table management and link status updates, scale less efficiently in dense mesh configurations compared to Thread’s more controlled and periodic control traffic dissemination.

When controlled application traffic is introduced, Matter over Thread maintains a low and relatively constant packet rate as the number of devices increases. Zigbee, by contrast, exhibits a pronounced growth in both total and application-layer traffic across all tested network sizes. This divergence becomes significantly more pronounced in the multi-hop chain topology, which places greater stress on routing and forwarding mechanisms.

As shown in Figure~\ref{fig:ChainLoad}, Matter over Thread demonstrates stable and approximately linear growth in total traffic as the hop count increases, indicating predictable scaling behavior. Zigbee, on the other hand, experiences a rapid escalation in traffic beyond three hops. This behavior is consistent with the operation of its reactive AODV-based routing protocol, which initiates network-wide broadcasts during route discovery. In low-bandwidth IEEE~802.15.4 networks, these broadcasts accumulate and lead to congestion effects that severely impact network stability.

In practice, this resulted in substantial degradation of Zigbee network reliability. During the five-hop experiment, only 94 out of 180 control commands were successfully delivered. In the six-hop configuration, the Zigbee coordinator terminated unexpectedly multiple times, preventing the collection of a stable data point. Matter over Thread remained fully operational under identical conditions, maintaining reliable command delivery across all tested network depths.

Similar congestion-induced phenomena in Zigbee networks have been reported in the literature. In \cite{Dinh2024WhenEvaluation}, excessive packet processing delays exceed the maximum end-to-end processing time of the Zigbee APS retransmission mechanism, leading to spurious application-layer retries. These retries further increase packet volume, compounding congestion and reinforcing the instability observed in multi-hop deployments.

 \begin{figure}[t]
     \centering
     \includegraphics[width=1\linewidth]{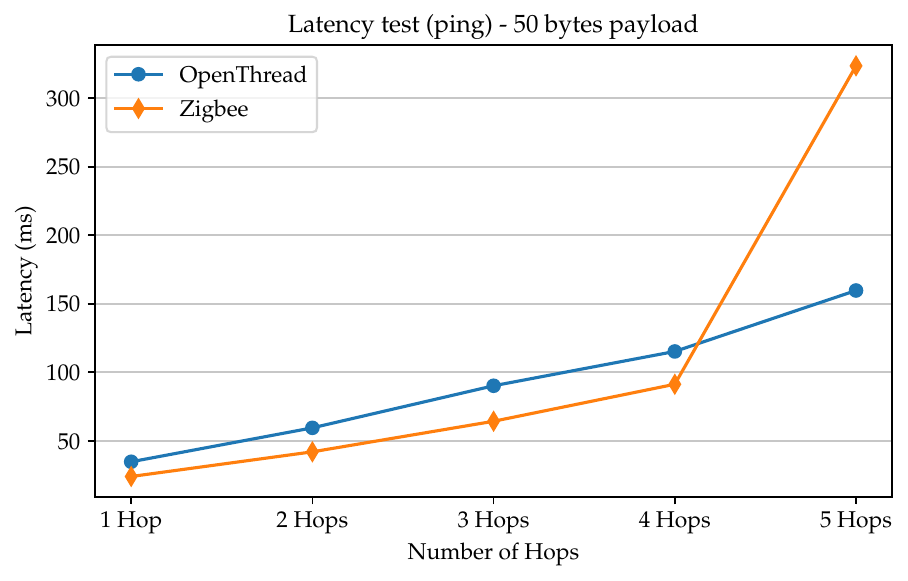}
     \caption{Latency Results over different number of hops (50-Byte payload)}
     \label{fig:Ping50ByHops}
 \end{figure}
 \begin{figure}[t]
     \centering
     \includegraphics[width=1\linewidth]{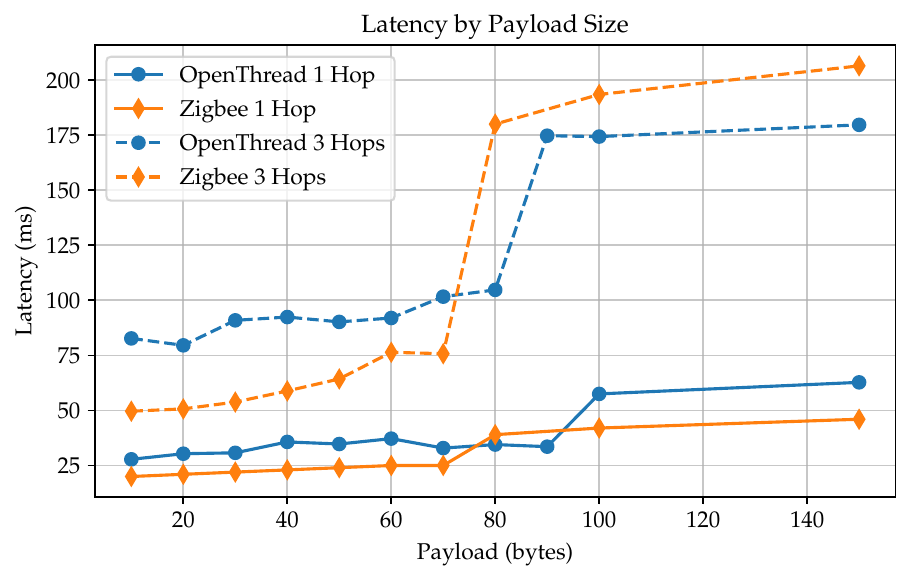}
     \caption{Latency results over different payload sizes (1 and 3 hops)}
     \label{fig:PingOTvsZBbyPayload}
 \end{figure}

\subsection{Latency and Throughput}
\label{sec:latency_throughput}

Following the overhead and scalability analysis presented in Section~\ref{sec:overhead}, we now examine how the observed protocol behaviors translate into end-to-end performance. In particular, this section evaluates the impact of routing, fragmentation, and congestion mechanisms on two fundamental network-level Key Performance Indicators (KPIs): end-to-end latency and maximum achievable throughput.

Unlike the overhead analysis, these experiments were designed to isolate the core transport- and network-layer performance of the two protocol stacks by explicitly excluding application-layer data processing. This approach enables a direct and equitable comparison of the underlying protocol mechanisms, unencumbered by the processing overhead of Matter or Zigbee application frameworks.

\subsubsection{Experimental methodology.}
Latency and throughput measurements were obtained using standard network benchmarking tools, specifically \texttt{ping} for round-trip time (RTT) measurements and \texttt{iperf} for throughput quantification. All experiments were conducted on ESP32-C6 development boards to ensure hardware consistency. The Zigbee testbed was based on the \texttt{esp\_zigbee\_all\_device\_types\_app} example provided by ESP-ZIGBEE-SDK version 1.6.5, while the Thread implementation relied on the \texttt{ot-cli} (OpenThread Command Line Interface) example from ESP-IDF version 5.4.1. Both firmwares expose a command-line interface via a serial-over-USB connection, enabling precise and repeatable control of the benchmarking tools.

To evaluate protocol performance across increasing network depths, a chain topology was strictly enforced using MAC address filtering. This configuration deterministically constrained the routing path and ensured that communication between the source and destination nodes necessarily traversed multiple hops. Experiments were repeated for network depths ranging from one to five hops.

\subsubsection{Latency measurements.}
End-to-end latency was quantified as the mean RTT obtained from successive \texttt{ping} exchanges between the controller and the destination node. Two complementary dimensions were explored: latency as a function of the number of hops for a fixed payload size, and latency as a function of payload size for fixed network depths.

Figure~\ref{fig:Ping50ByHops} reports the RTT measured with a fixed payload size of 50 bytes while increasing the number of hops. In single-hop configurations, Zigbee exhibits lower latency, achieving RTT values approximately 30\% lower than those observed for Thread. This behavior is attributable to the comparatively lightweight Zigbee protocol stack, which avoids the processing overhead associated with IP-based communication.

However, this advantage does not scale with network depth. As the number of hops increases, Zigbee latency grows rapidly and deviates from linear behavior. At higher hop counts, this increase is accompanied by significant packet loss, indicating congestion effects and reduced network stability. In contrast, Matter over Thread demonstrates a stable and predictable increase in RTT, scaling approximately linearly with the number of hops. This behavior reflects the more robust routing and forwarding mechanisms of Thread, which mitigate congestion and maintain consistent performance in multi-hop topologies.

The impact of payload size on latency is illustrated in Figure~\ref{fig:PingOTvsZBbyPayload}, which reports RTT measurements for increasing payload sizes under both single-hop and multi-hop conditions. For both protocols, latency increases once the payload exceeds the maximum frame size and fragmentation becomes necessary. However, Zigbee is more severely affected by fragmentation, particularly in multi-hop scenarios. In our experiments, fragmentation was observed for payload sizes exceeding approximately 79 bytes in Zigbee, compared to approximately 95 bytes for Thread in single-hop configurations and 89 bytes for Thread in networks with two or more hops. The comparatively better performance of Thread can be attributed to the presence of the 6LoWPAN adaptation layer, which provides more efficient and resilient fragmentation and reassembly mechanisms.

The RTT trends observed in these experiments are consistent with prior studies employing similar methodologies on different hardware platforms \cite{SiliconLabs2024AN1408:Performance,SiliconLabsAN1138:Performance}.

\subsubsection{Throughput measurements.}
Throughput performance was evaluated using the \texttt{iperf} tool by initiating a continuous data stream from the first node in the chain to the terminal node. Due to the architectural differences between the IP-based Thread protocol and the non-IP Zigbee stack, distinct optimization strategies were required.

For Thread, both User Datagram Protocol (UDP) and Transmission Control Protocol (TCP) transports were evaluated. To determine peak UDP throughput, a preliminary optimization phase explored different payload sizes, identifying approximately 800 bytes as the payload size yielding the highest throughput in multi-hop scenarios. While UDP achieved the highest absolute throughput, TCP-based measurements benefited from built-in flow control and congestion management, providing stable and repeatable performance without manual parameter tuning.

In contrast, Zigbee lacks native flow control mechanisms comparable to TCP. Achieving maximum stable throughput therefore required explicit regulation of the packet transmission rate to avoid congestion and packet loss. Through iterative empirical testing, the following optimal inter-packet intervals were identified: 7\,ms for 1 hop, 24\,ms for 2 hops, 42\,ms for 3 hops, 55\,ms for 4 hops, and 89\,ms for 5 hops. Notably, the 7\,ms interval identified for single-hop communication closely matches the theoretical transmission time reported in \cite{BurchfieldMaximizingImplementations}, providing validation for the experimental baseline.

\begin{figure}[t]
    \centering
    \includegraphics[width=1\linewidth]{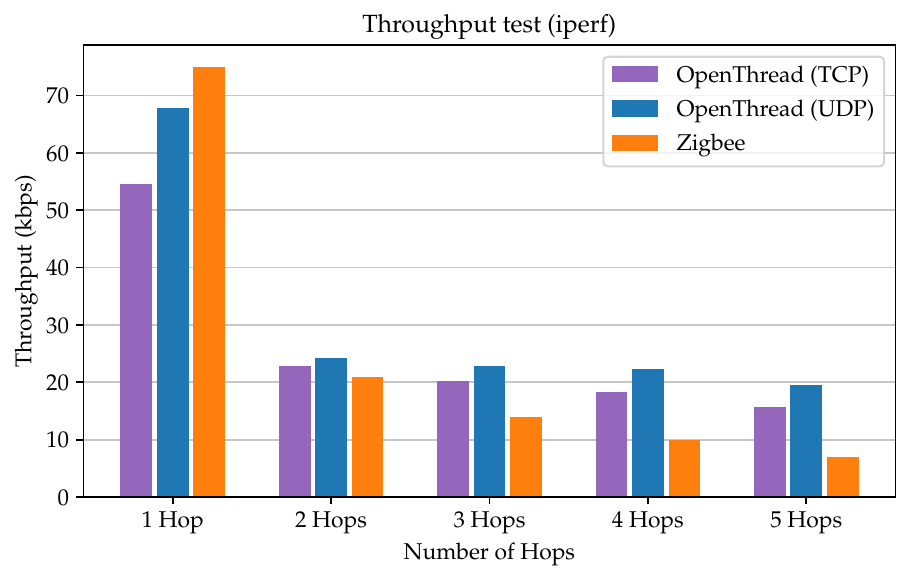}
    \caption{Throughput results as a function of the number of hops for Zigbee and Matter over Thread (UDP and TCP).}
    \label{fig:ThroughputByHops}
\end{figure}

The resulting throughput measurements are reported in Figure~\ref{fig:ThroughputByHops}. Zigbee achieves the highest peak throughput in direct single-hop configurations, reaching approximately 75\,kbps. However, consistent with the latency findings, its throughput degrades rapidly as additional hops are introduced. Matter over Thread, by contrast, maintains substantially higher throughput across multi-hop topologies. In particular, the TCP-based Thread configuration offers predictable and stable performance without requiring per-hop tuning, making it well suited for data-intensive operations such as Over-The-Air (OTA) firmware updates.

Overall, the throughput results observed in this study are comparable to, and in some cases exceed, those reported in prior industry evaluations \cite{SiliconLabs2019AN1142:Comparison}, likely reflecting differences in hardware platforms and protocol-level optimizations.

\subsection{Mesh Route Recovery}
\label{sec:recovery}

While the previous sections focused on steady-state performance metrics such as overhead, latency, and throughput, an equally critical property of mesh networking architectures is their ability to tolerate failures and recover from topology changes. In smart home deployments, where devices may become temporarily unavailable due to power loss or interference, fast and reliable route recovery is essential to maintain service continuity.

This section evaluates the route recovery behavior of Zigbee and Matter over Thread by quantifying the recovery latency following the failure of a critical intermediate router. Recovery latency is defined as the duration of the communication outage between the source and destination nodes.

\subsubsection{Experimental Methodology}

The experimental setup was designed to isolate network-layer recovery mechanisms from application-layer retry behaviors. The same hardware platform and firmware configurations used for the latency and throughput experiments were employed, ensuring consistency across all measurements. All over-the-air traffic was captured using a passive packet sniffer, enabling precise temporal analysis of the recovery process.

\begin{figure}[t]
    \centering
    \includegraphics[width=0.93\linewidth]{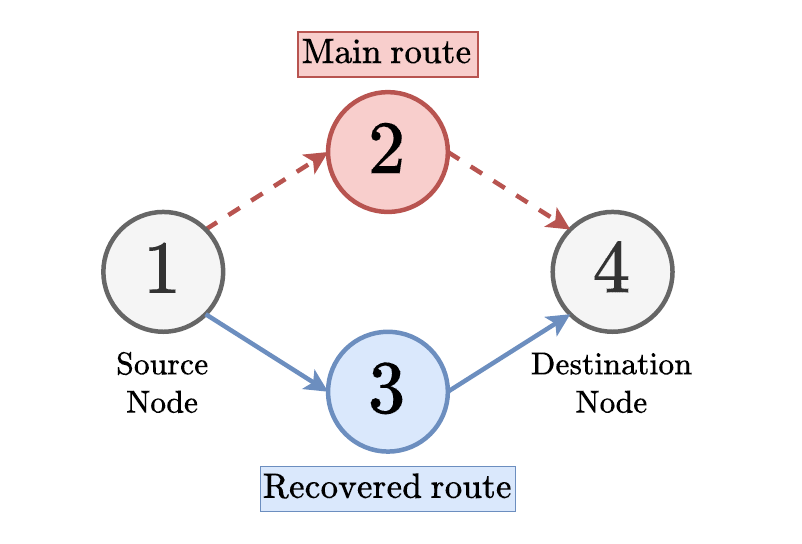}
    \caption{Topology used for the route recovery experiment, remembering the main route (via Node~2) and the recovered route (via Node~3).}
    \label{fig:RRtopo}
\end{figure}

A deterministic topology was constructed to ensure controlled and repeatable routing behavior. As illustrated in Figure~\ref{fig:RRtopo}, a four-node network was arranged in a diamond configuration, where communication between the source node and the destination node necessarily traverses one of two intermediate routers. This configuration provides exactly two redundant two-hop paths (e.g., Source–Node~2–Destination and Source–Node~3–Destination), guaranteeing the availability of a single, well-defined alternate route upon failure of the active path.

To trigger route recovery, a continuous stream of ICMP echo requests was generated to saturate the active route. Using the packet sniffer, the currently selected forwarding path was identified. The intermediate router responsible for relaying traffic on this active path was then abruptly powered off, emulating an ungraceful and unexpected node failure. The network’s subsequent behavior was monitored until packets were successfully delivered via the alternative route.

The Route Recovery Time was computed as the time difference between the timestamp of the last successfully transmitted packet on the failed route and the timestamp of the first successfully received packet on the recovered route. This destructive test was repeated 30 times for each protocol to ensure statistical validity.

\subsubsection{Results}

The experimental results reveal a marked divergence in the fault-tolerance strategies adopted by Zigbee and Matter over Thread, contrasting with the scalability trends observed in previous sections. The measured recovery times are summarized in Table~\ref{tab:RouteRecovery}.

Zigbee exhibits extremely fast recovery behavior, achieving an average route recovery time of 0.36\,s. This sub-second response reflects the reactive nature of Zigbee’s AODV-based routing protocol: upon detecting a broken link, the stack promptly initiates a Route Request (RREQ) broadcast to discover an alternative path. This behavior is consistent with prior experimental results reported in the literature, which document Zigbee recovery times on the order of 0.4\,s in comparable mesh configurations \cite{Ayurzana2024BuildingProtocol}.

In contrast, the Thread implementation demonstrates a significantly longer recovery phase, with an average communication outage of approximately 24\,s. Due to the limited availability of equivalent experimental results in the literature, this outcome was cross-validated using the OpenThread Network Simulator (OTNS), which yielded a closely matching recovery time. This confirms that the observed behavior is intrinsic to the protocol design rather than an artifact of the hardware platform.

The prolonged recovery time in Thread stems from its proactive routing architecture based on the Mesh Link Establishment (MLE) protocol. Unlike Zigbee’s reactive approach, Thread nodes rely on periodic routing advertisements and must wait for multiple missed announcements before declaring a neighbor unreachable. This design reflects an intentional trade-off that favors routing table stability and reduced control-plane overhead over rapid responsiveness to abrupt topology changes.

\renewcommand{\arraystretch}{1.5}
\begin{table}[t]
    \centering
    \caption{Route recovery time following an intermediate node failure}
    \begin{tabular}{|c|c|c|}
        \hline
        \textbf{Protocol} & \textbf{Average Time} & \textbf{Std. Dev.} \\ \hline
        OpenThread & 23.97\,s & 4.45\,s \\ \hline
        OpenThread (Sim.) & 24.45\,s & 3.71\,s \\ \hline
        Zigbee & 0.36\,s & 0.25\,s \\ \hline
        Zigbee (from \cite{Ayurzana2024BuildingProtocol}) & 0.40\,s & N/A \\ \hline
    \end{tabular}
    \label{tab:RouteRecovery}
\end{table}
\renewcommand{\arraystretch}{1}

\subsection{Discussion}
The analysis presented in this study demonstrates that neither Zigbee nor Matter over Thread exhibits universal superiority across all evaluated dimensions. Instead, the results reveal a fundamental trade-off between network agility and long-term operational stability, rooted in the contrasting routing and control-plane design choices of the two protocols.

Zigbee exhibits high agility, as evidenced by the latency and recovery experiments in Sections~\ref{sec:latency_throughput} and~\ref{sec:recovery}. Its lightweight protocol stack enables lower single-hop latency, while its reactive AODV-based routing allows rapid self-healing following intermediate node failures. These characteristics make Zigbee well suited for small to medium-scale deployments, such as residential smart homes, where network diameters are limited and rapid responsiveness is often prioritized. Additionally, Zigbee benefits from a mature ecosystem and broad device interoperability, which remain relevant factors in practical deployments. However, as shown in Section~\ref{sec:overhead}, Zigbee’s scalability is constrained by the control traffic generated by its reactive routing mechanisms. In larger or deeper mesh topologies, this overhead leads to congestion effects that negatively impact stability and reliability.

In contrast, Matter over Thread demonstrates superior stability and scalability under increasing network depth and load. The results in Sections~\ref{sec:overhead} and~\ref{sec:latency_throughput} show that Thread maintains predictable overhead growth, stable multi-hop latency, and substantially higher throughput in extended mesh configurations. These properties make Matter over Thread a more suitable foundation for large-scale and heterogeneous deployments, such as smart buildings or commercial environments. Its IP-based architecture and alignment with an emerging industry-wide standard further enhance its long-term viability. This stability, however, comes at the cost of reduced reactivity to topology changes. As demonstrated in Section~\ref{sec:recovery}, Thread’s proactive routing design leads to significantly longer route recovery times following node failures, which may be undesirable for applications with stringent availability or real-time requirements.

The obtained results indicate that protocol selection in IoT mesh networks should be driven by application-specific requirements rather than by a notion of absolute performance. Deployments prioritizing responsiveness and rapid fault recovery may benefit from Zigbee’s agile behavior, while scenarios emphasizing scalability, throughput, and long-term stability are better served by Matter over Thread. This trade-off is likely to remain central as IoT ecosystems continue to evolve toward larger, more heterogeneous, and increasingly mission-critical deployments.

\section{Conclusion}
\label{sec:conclusion}
This work presented a systematic experimental comparison of two widely adopted IoT communication protocols, Zigbee and Matter over Thread, with the goal of providing a data-driven assessment of their performance in realistic smart home mesh networks. Using a custom, protocol-agnostic testbed built on identical hardware platforms, we evaluated scalability, latency, throughput, and fault tolerance across multiple network topologies and traffic conditions.

The results highlight clear and consistent trade-offs between the two ecosystems. Zigbee demonstrates high agility, characterized by lower single-hop latency and rapid route recovery following node failures, making it well suited for small and relatively static deployments where responsiveness is critical. However, its reactive routing mechanisms lead to rapidly increasing control traffic and reduced stability as network size and depth grow. Matter over Thread, in contrast, exhibits predictable overhead scaling, stable multi-hop latency, and substantially higher throughput, establishing it as a more robust foundation for larger and more heterogeneous deployments. These advantages come at the cost of slower recovery from abrupt topology changes, reflecting a design emphasis on stability over reactivity.

Overall, the findings confirm that protocol selection in IoT mesh networks is inherently application-dependent and involves balancing agility, scalability, and long-term operational stability. Future work will extend this evaluation to larger-scale networks with higher node counts and deeper topologies, incorporate heterogeneous hardware platforms, and assess performance under realistic RF interference conditions. Additionally, a detailed analysis of energy consumption will be essential to complete the comparative evaluation of these protocols.
\balance

\section*{Acknowledgements}
This study was carried out within the MICS (Made in Italy
Circular and Sustainable) Extended Partnership and received
funding from Next-Generation EU (Italian PNRR M4 C2,
Invest 1.3 – D.D. 1551.11-10-2022, PE00000004). CUP MICS
D43C22003120001.

\bibliographystyle{IEEEtran}
\bibliography{bibliography}

\end{document}